\documentclass[%
 reprint,
 amsmath,amssymb,
 aps,
]{revtex4-1}

\usepackage{graphicx}
\usepackage{dcolumn}
\usepackage{bm}
\usepackage{xcolor}
\usepackage[normalem]{ulem}
\newcommand{\blue}[1]{\textcolor{black}{#1}}

\begin{document}

\title{Minigap and Andreev bound states in ballistic graphene}

\author
{L. Banszerus$^{1,2,*}$, F. Libisch$^3$, A.~Ceruti$^{1}$, S. Blien$^{4}$, K.~Watanabe$^5$, T.~Taniguchi$^6$, A.~K.~H\"uttel$^{4,7}$, B.~Beschoten$^{1}$, F. Hassler$^{8}$ and C. Stampfer$^{1,2}$ 
	\normalsize{\\$^1$JARA-FIT and 2nd Institute of Physics, RWTH Aachen University, 52074 Aachen, Germany, EU}\\
	\normalsize{$^2$Peter Gr\"unberg Institute (PGI-9), Forschungszentrum J\"ulich, 52425 J\"ulich, Germany, EU}\\
	\normalsize{$^3$Institute for Theoretical Physics, TU Wien, 1040 Vienna, Austria, EU}\\
	\normalsize{$^4$Institute for Experimental and Applied Physics, University of Regensburg, 93040 Regensburg, Germany, EU}\\
	\normalsize{$^5$Research Center for Functional Materials, 
		National Institute for Materials Science, 1-1 Namiki, Tsukuba 305-0044, Japan}\\
	\normalsize{$^6$International Center for Materials Nanoarchitectonics, 
		National Institute for Materials Science,  1-1 Namiki, Tsukuba 305-0044, Japan}\\
	\normalsize{$^7$Low Temperature Laboratory, Department of
		Applied Physics, Aalto University, Espoo, Finland, EU }\\
	\normalsize{$^8$JARA-Institute for Quantum Information, RWTH Aachen University, 52056 Aachen, Germany, EU}\\
	\normalsize{$^{*}$Corresponding author; E-mail: luca.banszerus@rwth-aachen.de}
}

\date{\today}

\begin{abstract}
A finite-size normal conductor, proximity-coupled to a superconductor has been predicted to exhibit a so-called minigap, in which quasiparticle excitations are prohibited. Here, we report on the direct observation of such a minigap in ballistic graphene, coupled to superconducting MoRe leads. The minigap is probed by finite bias spectroscopy through a weakly coupled junction in the graphene region and its value is given by the dimensions of the device. Besides the minigap, we observe a distinct peak in the differential resistance, which we attribute to weakly coupled Andreev bound states (ABS) located near the superconductor-graphene interface.
For weak magnetic fields, the phase accumulated in the normal-conducting region shifts the ABS in quantitative agreement with predictions from tight-binding calculations based on the Bogolioubov-de Gennes equation as well as with an analytical semiclassical model.  
\end{abstract}

\pacs{Valid PACS appear here}
\maketitle

Superconductivity is an intriguing 
quantum state of matter where electrons
pair up into Cooper pairs which in turn condense and form a collective
many-body ground state~\cite{Tinkham2004Jan}. Bringing a
superconductor (S) in electrical contact with a normal metal, semi-metal or semiconductor (N)
imprints some of the properties of superconductivity onto the
normal state, an effect that is called \emph{proximity effect} \cite{gennes}.
In a single particle picture, the proximity effect is due to electrons that,
originating from the normal region, impinge onto the superconductor~\cite{Pannetier2000}.  In this
process called Andreev reflection \cite{andreev:64} a hole is retroreflected as the electron has to form a Cooper pair
in order to be able to enter the superconducting condensate. The effect of
the superconductor is felt in a correlation of the electron and hole degrees of
freedom inside the normal conductor stretching over the coherence length $\xi = \hbar v_{\mathrm F} /\Delta$,
with $v_{\mathrm F}$ being the Fermi velocity in the N region and $\Delta$ being the gap of the bulk superconductor.
If the N region is smaller than the coherence length, the normal conductor behaves as a genuine superconductor,  i.e., it can carry a supercurrent and there is an energy range $2 \delta$ centered around the Fermi energy in which
there are no available states for quasiparticles~\cite{mcmillan:68}. 
This gap, called \emph{minigap} --- as $\delta$ is smaller than $\Delta$ --- can be estimated by 
$\delta \simeq \hbar/\tau$, where $\tau$ is the longest time between two
Andreev reflections~\cite{Belzig1999May}. 
For ballistic conductors in the regime $L<W<\ell_m$ (with length $L$, width $W$, mean-free path $\ell_m$), the minigap is inversely proportional to the width as the longest grazing trajectories cover the full width resulting in 
$\delta  \simeq \hbar v_{\mathrm F} / W = (\xi/W) \Delta$~\cite{gennes:63,golubov:88,belzig:96,pilgram:00,beenakker:05}.

\blue{Although spectral properties of various two-dimen-sional systems in the ballistic proximity regime have been probed by both transport measurements~\cite{Miao2007Sep,Mizuno2013Nov,Lee2015Jan,Ke2019Aug,BenShalom2015Dec} and tunneling spectroscopy~\cite{Dirks2011May,Bretheau2017Aug}, the proximity induced minigap has so far only been observed by tunneling experiments~\cite{Bretheau2017Aug}.} 
Good candidates for the search of minigap physics are tunable ballistic graphene-based Josephson junctions, which offer an intriguing platform to study the proximity effect, complex Andreev physics~\cite{BenShalom2015Dec,Park2019,Efetov2015,SanJose2015,Bretheau2017Aug,Nanda2017Jun,Calado2015Jul,BenShalom2015Dec,Seredinski2019Sep} and open the door to exotic topological phases 
in hybrid superconducting Dirac materials~\cite{SanJose2015}.
In this Letter, we present transport measurements on partially-gated ballistic graphene coupled to superconducting contacts, which allows for the direct observation of the minigap and magnetic field dependent Andreev bound states (ABS) with well-understood geometries whose energies cross the minigap. 
The spectral pattern of the ABS and the minigap are in good agreement with tight-binding calculations and a semiclassical model.
\begin{figure*}
    \centering
    \includegraphics[width=\linewidth]{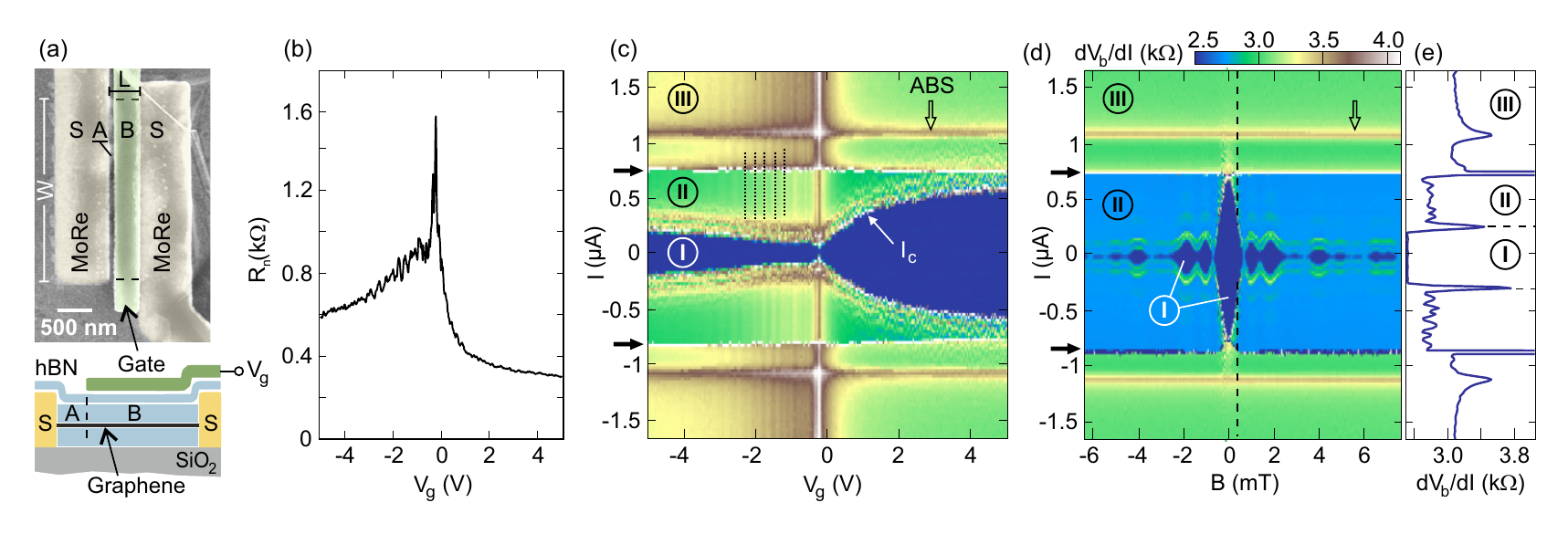}
    \caption{\textbf{(a)} 
    Top panel: False colored scanning electron microscope (SEM) image of the top-gated hBN/graphene/hBN device with MoRe contacts. Lower panel: Schematic illustration of the cross-section of the device highlighting the different graphene regions A and B.
    \textbf{(b)} Normal-state resistance $R_n$ as function of gate voltage measured at high bias ($V_b=7$\,mV, $T=15$\,mK). The parasitic line resistance $R_\mathrm{0}=2.55\,$k$\Omega$ has been subtracted. \blue{For negative gate voltages Fabry-P\'erot oscillations are observed}
    \textbf{(c)} 
    Differential resistance $\mathrm{d}V_b/\mathrm{d}I$ as function of applied current and gate voltage $V_g$, highlighting the different transport regimes I, II and III. The horizontal arrows mark the  minigap and the vertical arrow the ABS. \blue{The vertical dashed lines indicate Fabry-P\'erot oscillations} 
    \textbf{(d)} $\mathrm{d}V_b/\mathrm{d}I$ as function of current and small $B$-field ($V_g=5$\,V). \textbf{(d)} Line-trace from panel (d) for fixed $B$-fields ($B=0.3$\,mT, see dashed line in panel (d)).}
    \label{fig:exp_1}
\end{figure*}
The sample (see Fig. 1a) consists of a $L \approx 380$\,nm long and $W \approx 2.6\,\mu$m wide strip of 
graphene grown by chemical vapor deposition (CVD)~\cite{Banszerus2015,Banszerus2016Feb,Schmitz2020Sep}, encapsulated between two flakes of hexagonal boron nitride (hBN) and contacted by two sputtered superconducting MoRe electrodes \blue{(bulk gap $2\Delta\approx2.5\,$meV~\footnote{\blue{This energy has been estimated by $2 \Delta \approx 3.5 \; k_B T_c$ and using a critical temperature of MoRe of $T_c \approx 8.1$~K (see supplementary information). Here $k_B$ is the Boltzmann constant.}}).} 
After covering the structure by an additional hBN flake, a metallic top gate has been fabricated such that most of the graphene (region B) is covered by the gate, leaving an $\approx 60$\,nm strip of the graphene uncovered (region A). 
All measurements have been performed in a He$^3$/He$^4$ dilution refrigerator with a base temperature of 10\,mK. Crucially, the wiring and filtering of the setup imposes a constant parasitic resistance of $R_0 = 2.55$\,k$\Omega$ in series to the device, which has to be taken into account for finite bias measurements \blue{(see supporting material for details)}.

In Fig.~1b, we show the normal-state resistance $R_n$ as function of gate voltage $V_g$ that changes the carrier density $n_\mathrm{B}$ in the graphene region B (see cross-section in Fig.~1a). 
The charge neutrality point is shifted 
to negative voltages, which is due to electron doping induced by the sizeable work function mismatch between the MoRe contacts ($\approx 4$~eV~\cite{Babkin1974Nov}) and graphene ($\approx 4.5$ to $4.7$~eV).
As for ballistic graphene such doping is practically uniform away from the metal interface~\cite{Blake2009Jul}, the narrow, unscreened graphene region A is highly n-doped.

Depending on the current applied, the superconductor-graphene-superconductor (SGS) junction can be operated in different regimes.
This is highlighted in Fig.~1c, where 
we show the differential resistance $\mathrm{d}V_b/\mathrm{d}I$ as function of $V_g$ and dc current $I$.
As the expected coherence length $\xi = \hbar v_{\mathrm F} /\Delta \approx 480$\,nm ($v_{\mathrm F} = 10^6$\,m/s) is smaller than the length ($L$) of the graphene, the SGS junction exhibits for $I < I_c$ a fully developed proximity effect (regime I). Here, the critical current $I_c$ increases with $|V_g|$ in good agreement with the decrease of $R_n$ (Fig.~1b).

As in Ref.~\cite{BenShalom2015Dec}, we observe for negative $V_g$  Fabry-P\'erot oscillations, both in the normal state as well as in $I_c$ (see vertical dashed lines in Fig.~1c \blue{and Fig.~S1 of the supporting material}), which are in agreement with the periodicity given by the 320\,nm cavity-length of region B, where the pn-junction at the A-B interface serves as one of the mirrors.
\blue{For currents above $I_c$ the device is turned into an effective SN junction, where a finite voltage drops over the A-B interface and no supercurrent can be observed anymore. 
As $I_c$ is mainly limited by the normal state resistance of region B ($|n_\mathrm{B}| \ll |n_\mathrm{A}|$, $I_c$ remains gate-dependent) for currents only slightly exceeding $I_c$, the voltage drops over region B, while region A remains superconducting (regime II).}
As the device is ballistic, the transport is dominated by Andreev processes involving semiclassical trajectories that run perpendicular to the interfaces. These trajectories (originating from the B region), cross the A-B interface with almost unit probability before being retroreflected at the A-S interface via an Andreev reflection. This allows to probe the proximity-induced minigap in region A appearing as a pronounced feature around $I \approx 0.75~\mu$A in Figs.~1c and 1d (see horizontal arrows). 
The reduction of the Andreev conductance at this point is due to the fact that these additional states open a new scattering channel for the perpendicular trajectories dominating the transport. In this sense, the differential resistance serves as a probe of the density of states in the region A.
By further increasing the current, we excite the system above the minigap (regime III) and observe a feature of enhanced differential resistance, which we attribute to a weakly coupled ABS  (see vertical arrows in Figs.~1c,d). 
In contrast to the critical current $I_c$, both the signatures of the minigap and the ABS show only a very weak dependence on
which implies that they are located in region A. 
\begin{figure*}[t]
    \centering
    \includegraphics[width=\linewidth]{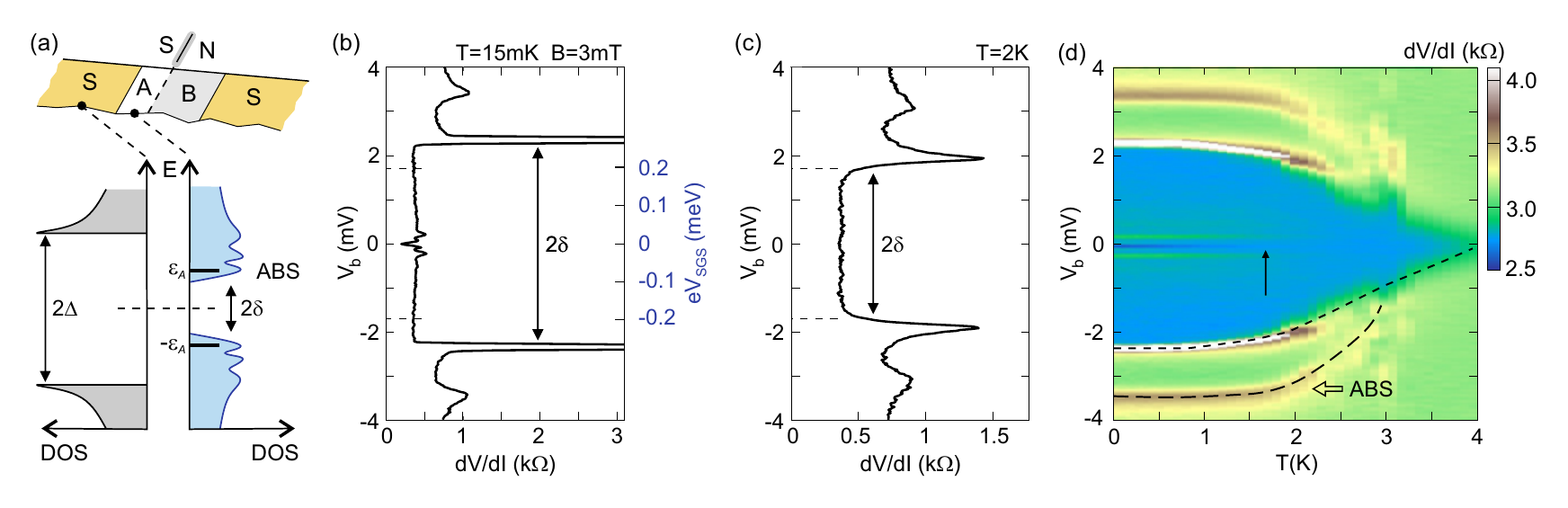}
    \caption{\textbf{(a)}
    Schematics of the density of states (DOS) in the different regions of the sample when forming an effective SN-junction, where region A is proximitized, highlighted by the minigap (2$\delta$) and  Andreev bound states (ABS). \textbf{(b)} Corrected differential conductance, $dV/dI$ as function of bias voltage for $B=3$\,mT and $T=15$\,mK. The constant differential resistance at low bias allows to include the energy scale on the right $y$-axis. \textbf{(c)} Similar measurement as in panel b, recorded at $T=2$\,K. \textbf{(d)} $dV/dI$ as function of $V_\mathrm{b}$ and temperature. The dashed lines mark the structures connected to the minigap and ABS. } 
    \label{fig:exp_2}
\end{figure*}
By applying a small out-of-plane magnetic field $B$ for fixed $V_g=5$\,V, we observe a Fraunhofer interference pattern~\cite{Tinkham2004Jan}. The deviations from the typical sinc($B$) behaviour occurs due to  the non-trivial current-phase relation in graphene~\cite{Nanda2017Jun}, the missing translational invariance of the device along the transport direction, and potential flux trapping in the MoRe leads. 
Note that the positions of the structures related to the minigap and the ABS remain constant for these small $B$-fields. Figure~1e shows a line cut of Fig.~1d at $B=0.3$\,mT highlighting the sharpness of the  peaks separating the different transport regimes (see labels). 

Next we focus on the energy scale of the observed minigap. 
When operating the device slightly above $I_c$, we have an effective SN junction that allows via the measurement of the differential conductance to probe the density of states (DOS), including the minigap of the partly proximitized graphene (region A only, see illustration in Fig.~2a).
Note that the minigap is the same everywhere in the proximitized structure~\cite{leSueur2008May}, i.e. in the A region, which implies a non-zero local DOS between $\delta$ and $\Delta$, which hosts Andreev states being reflected at the interface of the S and A region.

In Fig.~2b, we show the corrected differential resistance $dV/dI=dV_\mathrm{b}/dI-R_0$ as function of $V_\mathrm{b}$. As these data are taken at a small finite $B$-field ($B \approx 3$\,mT), the proximity effect and thus $I_c$ is strongly suppressed and only features related to the minigap and the ABS remain visible. As $dV/dI$ is almost constant for small bias voltages, we can determine the fraction of the applied bias voltage that drops over the SGS junction, i.e, the effective SN junction by $V_\mathrm{SGS}=V_\mathrm{b} (dV/dI)/(dV_\mathrm{b}/dI)$.
The energy scale
$e V_{\mathrm{SGS}}$ finally allows to extract
the energy of the minigap resulting in $2 \delta \approx 0.5$~meV (see arrows in Figs.~2b,c). This value is in excellent agreement with what we expect when using 
$\delta  \simeq \hbar v_{\mathrm F} / \ell_m$
with $\ell_m \simeq W$ (ballistic regime)
leading to $\delta \approx 0.25$\,meV, confirming the detection of the minigap in the ballistic region A. \blue{Note that the features related to the bulk gap of the MoRe electrodes are expected at much higher energies close to $eV_\mathrm{SGS}=\Delta\approx1.25$~meV.}

In Fig.~2c, we show a similar measurement as in Fig.~2b but for $T=2$\,K and $B \approx 3$\,mT highlighting that both, the peak attributed to the minigap and the ABS smear out but survive. 
Indeed, the signatures of the minigap and the ABS are nearly unchanged up to temperatures of 1.5\,K, as shown in Fig~2d, indicating that both are determined by semiclassical trajectories depending only on the geometry in agreement with being ballistic. Interestingly, the differential resistance peak associated with the ABS starts then to shift to lower energies and rapidly vanishes in the range between 2 and 3\,K. 
This is in contrast to what we observe for the characteristics of the minigap where the region of the reduced $dV/dI$ decreases linearly for
$\approx 2$\,K and disappears only at $\approx 4$\,K. 
In both cases we assume that for temperatures $>1.5$~K inelastic scattering processes set in, which lead to dephasing and thus suppress the presence of the ABS and the minigap.
As the ABS are based on longer trajectory the impact is more dramatic and the related feature vanishes at lower temperate compared to the minigap, demonstrating their different nature.

Now we turn to the ABS and discuss their magnetic field dependence, which allows (i) for a comparison with theory and, more importantly, (ii) for a geometrical interpretation of the underlying ballistic electron and hole trajectories (see Fig.~3a).
In Fig.~3b, we show that for larger magnetic fields the energy of the ABS $\varepsilon_A$ decreases well below $\delta$ and then sharply moves out of the minigap again forming triangular features at around $B\approx150$\,mT. Here, the fully proximitized regime is not visible due to its strong suppression due to flux trapping at high magnetic fields as well as due to measurement resolution (compare $B$-field scales in Fig.~1d and Fig.~3b).

To elucidate the origin of the $B$-field dependence of the observed resistance peak (i.e., the weakly coupled ABS) we simulate a graphene SN junction. 
We consider a 2\,$\mu$m wide graphene stripe, which is superconducting for $x < 0$,
  normal conducting for $x > 0$ and features open boundary conditions on both sides ($x<0$ and $x > 100$\,nm) to 
  describe an extended quantum system. Our simulation is based on a third-nearest neighbor 
  tight binding description of graphene using the Bogolioubov-de Gennes equation~\cite{Beenakker2006Aug},
  \begin{equation}\label{eq:BdG}
      \left(\begin{array}{cc}
          H - E_F& \Delta(x) \\ \Delta^*(x) & E_F - H^*
      \end{array}\right)      \left(\begin{array}{cc}
          \phi_e\\ \phi_h
      \end{array}\right) = \varepsilon     \left(\begin{array}{cc}
          \phi_e\\ \phi_h
      \end{array}\right),
  \end{equation}
where we choose $E_F = 170$\,meV (electrons) in the region without top gate (region A), and $-30$ meV (holes) under the top gate (region B). We find the same qualitative behavior
for different potential configurations, as long as $|n_\mathrm{B}| \ll |n_\mathrm{A}|$. To make the size of our system manageable, we rescale spatial coordinates
by a factor of ten while keeping the band structure the same. Inside the superconductor, $\Delta(x) = \Delta_0 = 1.2$\,meV. In the normal conducting region, we include the magnetic field by a Peierl's phase, while keeping $B=0$ in the superconducting part of the sample due to the Meissner effect. We include a randomly correlated disorder potential with an amplitude $\sqrt{\langle V^2\rangle} = 20$\,meV ($\langle V\rangle = 0$) and correlation length of 20\,nm to average over details of the potential landscape.
\begin{figure}[t]
    \centering
    \includegraphics[width=\linewidth]{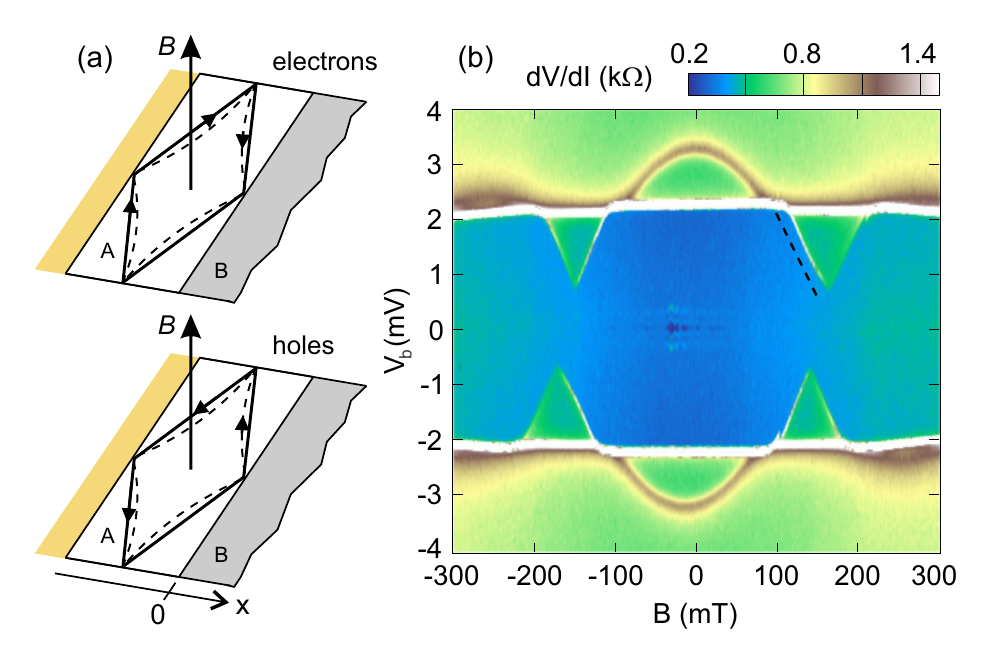}
    \caption{\textbf{(a)} Schematic illustrations of the semiclassical electron (upper panel) and hole (lower panel) trajectories of the Andreev bound states used to analytically estimate the $B$-field dependence of $\varepsilon_\mathrm{A}$. The enclosed area is about half the area of region A. The dashed lines show the trajectories for finite $B$-field. \textbf{(b)} $dV/dI$ as function of $V_\mathrm{b}$ and magnetic fields between $\pm300$~mT. The dashed line indicates the slope of the semiclassical model (see text).}
    \label{fig:exp_3}
\end{figure}
\begin{figure}[t]   
\centering
    \includegraphics[width=0.9\linewidth]{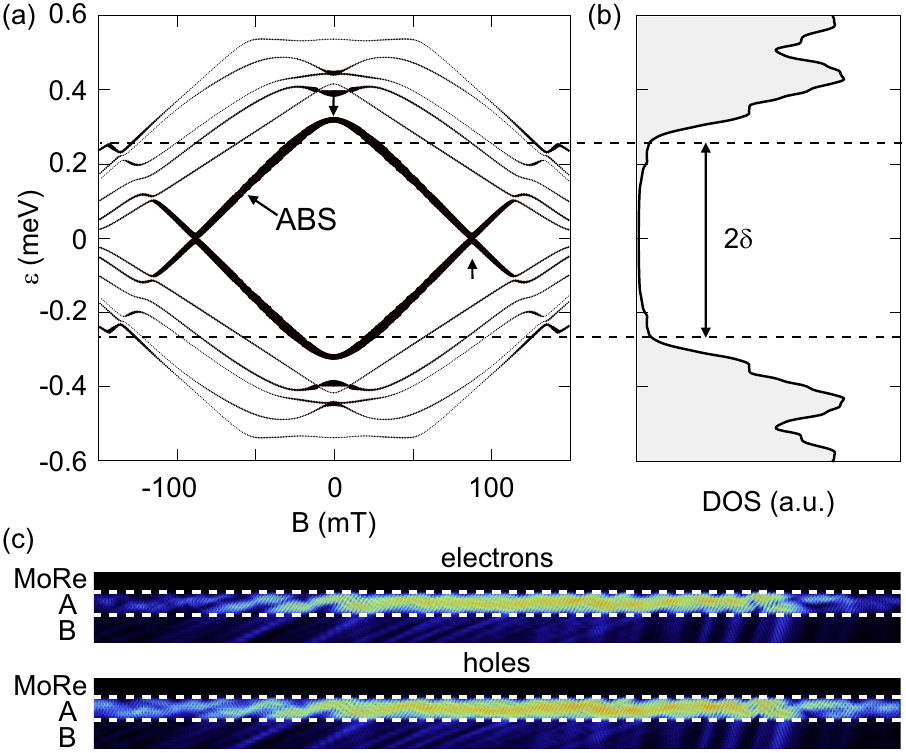}
    \caption{\textbf{(a)} Eigenenergies of one realization of the model Andreev system (see text) as function of $B$-field. Line thickness is given by the inverse coupling strength, $\Gamma$, to region B. \textbf{(b)} Disorder-averaged DOS of the model Andreev system. Dashed lines mark the minigap as guide to the eye in (a) and (b). \textbf{(c)} Electron and hole wave function of
    the lowest-energy eigenstate (marked ABS in (a)) at $B=0$. Notice the strong localization of the eigenstate within the A region and the weak coupling to region B.}
    \label{fig:theory_wavef}
\end{figure}
We calculate eigenstates of the cavity around $\varepsilon=0$ (Fig.~4a), and find a sizeable minigap $\delta$ (Fig.~4b): Due to the disorder, the system becomes chaotic, introducing a gapped region around the Fermi energy. We find a minigap of $\delta \approx 0.25$\,meV, well in line with our experimental estimates above, and substantially smaller than the superconducting gap, $\delta \ll\Delta_0$. Since we solve an open quantum system, we obtain complex eigenenergies $E = \varepsilon + i \Gamma$ whose imaginary part $\Gamma$ is related to the coupling to the open boundaries. We find the smallest $\Gamma$ (of the order of $0.1$\,$\mu$eV) for the lowest-lying state, while delocalized Bloch states feature larger couplings to the leads. 
The wavefunction of this state features an orbit propagating parallel to the SN interface (see Fig.~\ref{fig:theory_wavef}c), and consequently interacts only weakly with the top-gated B region, resulting in sharp features in the density of states. Including a magnetic
field causes the energy of the ABS to decrease (Fig.~\ref{fig:theory_wavef}a), in quantitative correspondence with experiment. The magnetic-field evolution of the state energy $\varepsilon_A(B)$ can be fitted almost perfectly using
\begin{equation}\label{eq:pertfit}
\varepsilon_A = \varepsilon_{A,0} - \sqrt{\gamma^2 + \alpha^2 B^2},
\end{equation}
reminiscent of degenerate perturbation theory with a zero-field
splitting $\gamma$ given by the disorder. For magnetic fields $\alpha B \gg \gamma$
the magnetic field dependence of $\varepsilon_A(B)$ becomes approximately linear,
$\varepsilon_A(B) \approx \varepsilon_0 - \alpha B$, with a slope of $-\alpha$.
To obtain an analytical estimate for $\alpha$, we consider a semiclassical picture \cite{PRB66CBKV} where the ABS correspond to the orbit 
sketched in Fig.~3a, enclosing an area $a$ of
about half of region A. Such an orbit, indeed, features a  grazing angle
of incidence on the A--B interface, and thus a small transmission (which explains the weak coupling to the outside and thus the sharpness of the peak in the data). According to a simple
semiclassical Bohr-Sommerfeld quantization, the action along the semiclassical path needs to equal some constant phase $\phi_C$ given by the boundary conditions
\cite{PRB66CBKV},
\begin{equation}
     \frac{\varepsilon_A \ell_A}{\hbar v_F} +  \frac{a B}{\Phi_0} = \phi_C,  
\end{equation}
where the second term is due to the enclosed magnetic flux. The value of $\phi_C$ drops out when determining $\alpha$ as we bring the above equation into the form
$\varepsilon_A = \mathrm{const}\cdot\phi_C -\alpha B$. 
Inserting the geometry at hand, with $a \approx \tfrac12  (2.6\,\mu$m${}\times60$\,nm) and $\ell_A \approx 2W = 5.2\,\mu$m yields $\alpha_{\mathrm{sem}} = a \hbar v_F / (\ell_A\Phi_0) \approx 4.2$\,meV/T, in close 
 agreement with an average over 145 disorder realizations for our numerical model, 
$\alpha_{\mathrm{num}} \approx (4 \pm 1) $\,meV/T and also with the experimental value
 $\alpha_{\mathrm{exp}}=(3.3\pm0.4)$\,meV/T (extracted from the dashed line in Fig.~3b~\footnote{The error comes from averaging over the four slopes (positive/negative $V_b$ and positive/negative $B$-field).}).

For larger magnetic fields, our calculations do not reproduce the sharp, teeth-like features and oscillating patterns of experiment. Instead, we find multiple avoided crossings between states converging towards $\varepsilon_A/\Delta = 0$. 
\blue{We conjecture that the experimental features are related to vortices that can tunnel in the type-II superconductor, which are not included in our model. They would cause the local magnetic field to deviate from the external one --- as corroborated by the hysteresis and asymmetry in $B$ field observed in experiment. Furthermore, the penetration depth is dependent on the applied magnetic field, which has an influence on the trajectory lengths. To describe these effects at high magnetic fields, a self-consistent calculation would be required.}

In conclusion, we have shown the first observation of a minigap in a ballistic system by transport experiments. The minigap arises to due proximity coupling of \blue{ballistic} graphene to superconducting leads. We can reproduce the evolution of the Andreev bound states with a full simulation of the experimental geometry as well as with a semiclassical model. 
Using a model system based on the Bogolioubov-de Gennes equation and experimental parameters such as the device geometry we can reproduce
ABS that couple weakly to the normal part of the graphene system. The magnetic field dependence of its energy agrees well with
experiment, and with an analytical estimate based on Andreev orbits. 
Our work demonstrates that graphene offers a highly interesting playground for testing theoretical predictions made for Andreev billards~\cite{beenakker:05}.
This promise is extended by recent developments in bilayer graphene, where electrostatically defined cavities have  been reported, as well as the detection of supercurrents in the quantum Hall regime~\cite{Amet2016May}  opening the door to search for exotic topological phases, including Majorana zero modes~\cite{SanJose2015}.
\newline
\newline
{\bf Acknowledgment --- } We thank U.~Wichmann for help with the measurement electronics and  \.{I}.~Adagideli and W.~Belzig for stimulating discussions. This project has received funding from the Deutsche Forschungsgemeinschaft (DFG, German Research Foundation) under Germany's Excellence Strategy – Cluster of Excellence Matter and Light for Quantum Computing (ML4Q) EXC 2004/1 -- 390534769, the European Union's Horizon 2020 research and innovation programme under grant agreement No. 881603 (Graphene Flagship) and from the European Research Council (ERC) (grant agreement No. 820254)), and the Helmholtz Nano Facility~\cite{Albrecht2017May}. K.W. and T.T. acknowledge support from the Elemental Strategy  Initiative  conducted  by  the  MEXT, Japan, Grant Number JPMXP0112101001, JSPS KAKENHI Grant Number JP20H00354 and the CREST(JPMJCR15F3), JST.

\bibliographystyle{apsrev4-1}

\bibliography{literature}

\end{document}


\title{Supporting information for: ~\\ Minigap and Andreev bound states in ballistic graphene}

\author
{L. Banszerus$^{1,2,*}$, F. Libisch$^3$, A.~Ceruti$^{1}$, S. Blien$^{4}$, K.~Watanabe$^5$, T.~Taniguchi$^6$, A.~K.~H\"uttel$^{4,7}$, B.~Beschoten$^{1}$, F. Hassler$^{8}$ and C. Stampfer$^{1,2}$ 
	\normalsize{\\$^1$JARA-FIT and 2nd Institute of Physics, RWTH Aachen University, 52074 Aachen, Germany, EU}\\
	\normalsize{$^2$Peter Gr\"unberg Institute (PGI-9), Forschungszentrum J\"ulich, 52425 J\"ulich, Germany, EU}\\
	\normalsize{$^3$Institute for Theoretical Physics, TU Wien, 1040 Vienna, Austria, EU}\\
	\normalsize{$^4$Institute for Experimental and Applied Physics, University of Regensburg, 93040 Regensburg, Germany, EU}\\
	\normalsize{$^5$Research Center for Functional Materials, 
		National Institute for Materials Science, 1-1 Namiki, Tsukuba 305-0044, Japan}\\
	\normalsize{$^6$International Center for Materials Nanoarchitectonics, 
		National Institute for Materials Science,  1-1 Namiki, Tsukuba 305-0044, Japan}\\
	\normalsize{$^7$Low Temperature Laboratory, Department of
		Applied Physics, Aalto University, Espoo, Finland, EU }\\
	\normalsize{$^8$JARA-Institute for Quantum Information, RWTH Aachen University, 52056 Aachen, Germany, EU}\\
	\normalsize{$^{*}$Corresponding author; E-mail: luca.banszerus@rwth-aachen.de}
}

\date{\today}
\maketitle

\begin{figure}[h]
    \centering
    \includegraphics[width=0.5\linewidth]{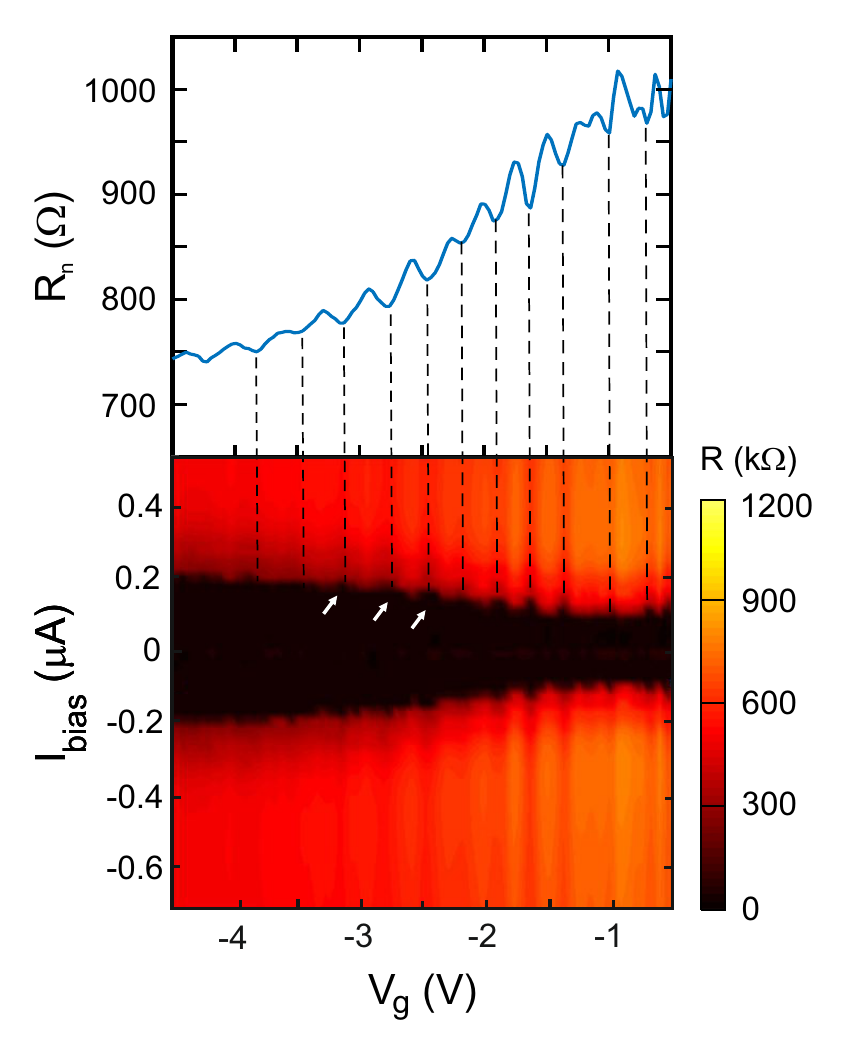}
\caption[FigS1]{Zoom-in on the Fabry-P\'erot oscillations observed at negative gate voltage. Upper panel: Normal resistance showing clear Fabry-P\'erot resonances. Lower panel: Resistance as function of $I$ and $V_g$. The critical current is modulated with the Fabry-P\'erot oscillations (see white arrows).}
\end{figure}

\begin{figure}[t]
    \centering
    \includegraphics[width=0.85\linewidth]{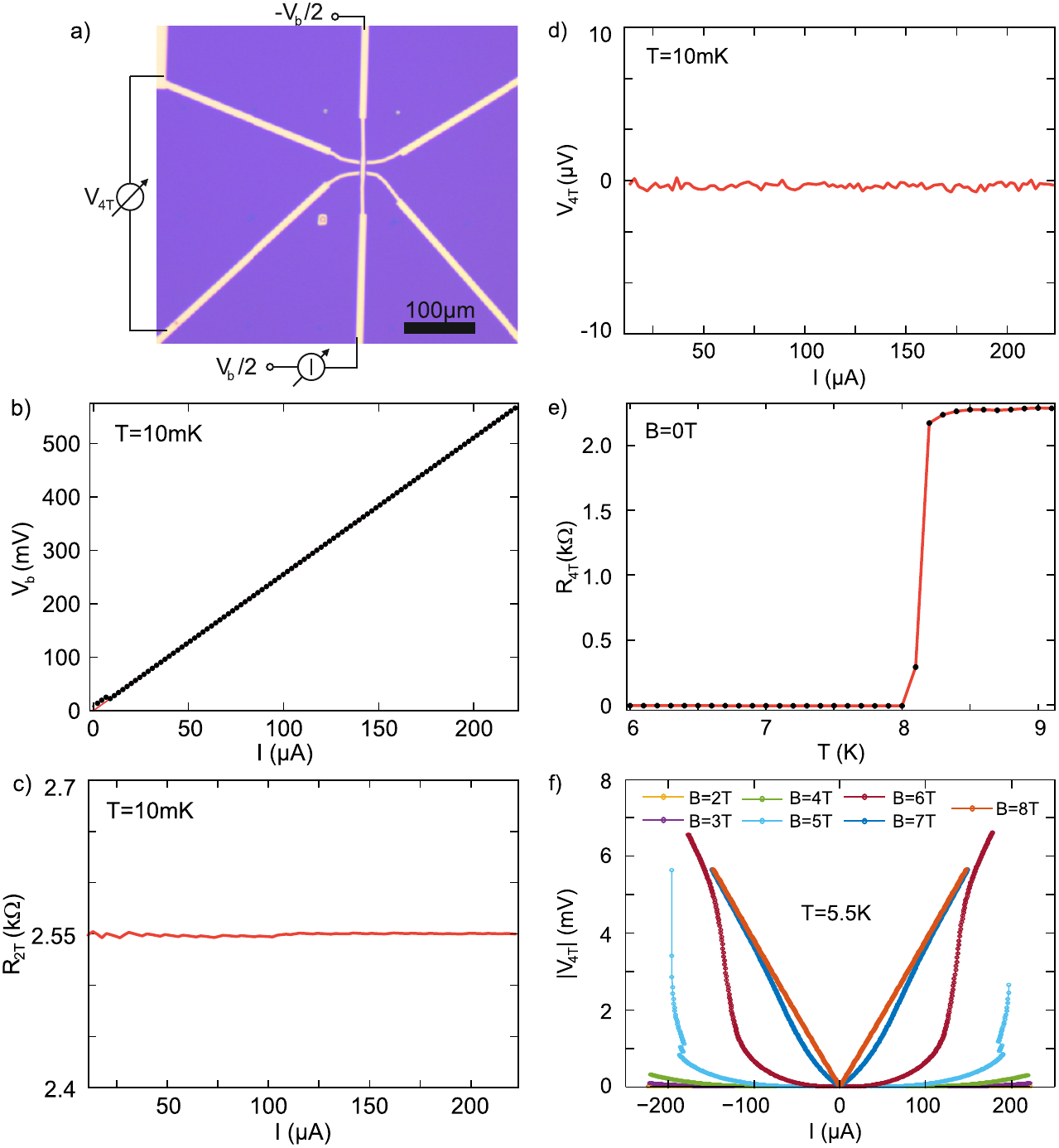}
    \caption[FigS2]{\textbf{(a)} Optical microscope image of a MoRe reference structure fabricated in the same sputtering run as the MoRe/graphene/MoRe device discussed in the main text. The data in \textbf{(b)-(f)} have been measured in the very same setup as the device discussed in the main text using the identical lines. \textbf{(b)} I-V characteristic of the reference device measured at 10~mK. The I-V trace shows a perfectly linear behaviour without subgap features. The slope allows to extract a line resistance of $R_0=2550~\Omega$ with an uncertainty of $10~\Omega$ (see panel~\textbf{(c)}) and no voltage drop is observed in the four terminal measurement (see panel \textbf{(d)}) indicating a clean superconducting behaviour of the bulk MoRe. \textbf{(e)} Temperature dependence of the two terminal resistance, showing a critical temperature of around $T=8$~K. \textbf{(f)} Four terminal voltage drop as function of current, recorded at $T=5.5$~K for various magnetic fields. A transition to the normal state is observed for fields exceeding $B=4$~T. At 10~mK, $H_{c2}$ exceeds 8~T and could not be experimentally observed.}
    \label{MoRe}
\end{figure}

\bibliographystyle{apsrev4-1}

\bibliography{literature}